\documentclass[a4paper,11pt]{article}
\pdfoutput=1 
\usepackage{jcappub}

\usepackage[english]{babel}
\usepackage[utf8]{inputenc}

\usepackage{graphicx}
\usepackage{amsmath}
\usepackage{amssymb}

\usepackage{hyperref}

\title{Probing Velocity Dependent Self-Interacting Dark Matter with
  Neutrino Telescopes}
\author{Denis~S.~Robertson}
\author{and Ivone~F.~M.~Albuquerque}
\affiliation{Instituto de Física,\\Universidade de São Paulo, Brazil}

\emailAdd{denistefanrs@gmail.com}
\emailAdd{ifreire@if.usp.br}

\abstract{Self-interacting dark matter models constitute an attractive solution to problems in 
structure formation on small scales. A simple realization of these models considers the
dark force mediated by a light particle which can couple to the Standard Model through
mixings with the photon or the $Z$ boson. Within this scenario we investigate the 
sensitivity of the IceCube-DeepCore and PINGU neutrino telescopes to the associated muon 
neutrino flux produced by dark matter annihilations in the Sun. Despite the model's 
simplicity, several effects naturally appear:
momentum suppressed capture by nuclei, velocity dependent dark matter self-capture, Sommerfeld 
enhanced annihilation, as well as the enhancement on the neutrino flux due to mediator 
late decays. Taking all these effects into account, we find that most of the model relevant parameter space can be tested by 
the three years of data already collected by the IceCube-DeepCore. We show that indirect 
detection through neutrinos can compete with the strong existing limits from direct 
detection experiments, specially in the case of isospin violation.
}
\arxivnumber{1234.56789}

\begin{document}

\maketitle

\section{Introduction}


Cosmological observations  position $\Lambda$CDM as the standard model of 
cosmology. In this picture the Universe is dominated by dark energy 
(consistent with a cosmological constant $\Lambda$) and collisionless cold dark 
matter (CDM). Despite its great success at large scales 
($\vartheta(\mathrm{\gtrsim Mpc}))$, 
the CDM hypothesis encounters some potential difficulties describing 
smaller scales \cite{Weinberg:2013aya,DelPopolo:2016emo}. These arise since 
observations do not agree with predictions from structure formation simulations. One 
of these problems is the core-cusp \cite{deBlok:2009sp}  
discrepancy between the profiles of dark matter halos 
observed in dwarf and low surface brightness galaxies, which present a flat density 
core \cite{Oh:2010ea,KuziodeNaray:2007qi} and the density profiles found in high 
resolution N-body CDM simulations, which steeply grow toward the center 
\cite{Navarro:1996gj,Navarro:2008kc,Stadel:2008pn}. 
More recently the so called too big to fail (TBTF) problem 
\cite{BoylanKolchin:2011de,BoylanKolchin:2011dk} came into evidence, and refers 
to the fact that $\Lambda$CDM simulations for Milky Way--like galaxies predict a number of massive 
subhalos that are too dense to be consistent with our observed galaxy's dwarf 
satellites.


A promising possibility to solve or at least alleviate these problems is self-interacting 
dark matter (SIDM) \cite{Spergel:1999mh}. Within this scenario dark matter particles 
interact with each other, allowing energy transfer from the external hotter regions of 
a dark matter halo to its cooler center. As a consequence the halo central density 
decreases forming a cored profile. Several SIDM simulations support this hypothesis, 
leading to a better agreement with observations when compared to CDM simulations 
\cite{Burkert:2000di,Moore:2000fp,Dave:2000ar}. To alleviate the TBTF problem, it is 
shown that the ratio between the dark matter self-interaction cross section and its 
mass must be within $0.1 - 0.5 \lesssim \sigma/m_\chi \lesssim 10 - 50$ cm$^2$/g 
\cite{Vogelsberger:2012ku,Zavala:2012us,Elbert:2014bma}. On the observational side 
there are very strong constraints on $\sigma/m_\chi$. Among these, the most stringent 
comes from analysis of the Bullet Cluster \cite{Randall:2007ph,Robertson:2016xjh} and 
other colliding clusters \cite{Harvey:2015hha}. There are also more model dependent 
constraints as the ones from indirect dark matter probes through neutrino detection 
\cite{Albuquerque:2013xna}, and from simulations of structure formation 
\cite{Vogelsberger:2012ku,Zavala:2012us,Elbert:2014bma}, and based on dark matter halo
shapes and density profiles of galaxies and clusters \cite{Rocha:2012jg,Peter:2012jh}.
All these combined require that $0.1 \lesssim \sigma/m_\chi \lesssim 0.5$ cm$^2$/g, 
and thus exclude a large fraction of SIDM as a solution to the CDM small scale 
problems.


Nevertheless, these observations are reconciled if the dark matter self-interacting
cross section depends on the particles' relative velocity. 
These scenarios (vdSIDM), occur naturally in SIDM models containing a dark force described by a Yukawa potential 
\cite{Feng:2009hw,Loeb:2010gj,Tulin:2013teo}. In these simple models the cross section 
is large at low velocities and falls rapidly as the relative velocity
increases, so they alleviate the TBTF problem of dwarf satellites
whose velocities are characteristically low (around $\sim 10$ km/s). 
At the same time, vdSIDM also evades the strong constraints from larger 
astrophysical systems mentioned above (which have characteristic velocities
between $\sim 200 - 4000$ km/s).

In spite of its simplicity, vdSIDM models have a rich phenomenology. 
As an example, besides their non-trivial self-scattering cross section \cite{Tulin:2013teo}, 
their annihilation is enhanced by the Sommerfeld effect \cite{ArkaniHamed:2008qn,Feng:2010zp}.
When mixing between the dark force mediator and the Standard Model (SM) is considered, 
dark matter particles can scatter with nucleons, which in conjunction with the fact 
that their annihilation produces 
SM particles, opens up many possibilities for vdSIDM detection.
Possible couplings are kinetic mixing with the photon or mass mixing
through the $Z$ or the Higgs boson.
Several of these models have been probed, either 
by direct detection \cite{Kaplinghat:2013yxa,DelNobile:2015uua} or indirect detection 
techniques \cite{Bringmann:2016din,Cirelli:2016rnw}, imposing quite strong constraints 
on specific vdSIDM models. 


In this work we  explore how vdSIDM modifies the high energy neutrino flux from dark 
matter annihilation in the Sun, extending our work which indirectly constrains non 
velocity dependent SIDM \cite{Albuquerque:2013xna}. We estimate the vdSIDM annihilation 
neutrino rate, and by comparing it to the atmospheric neutrino background 
IceCube-DeepCore's and PINGU's neutrino telescopes' sensitivity to vdSIDM. We 
therefore provide grounds for an independent and complementary probe on vdSIDM.

Our analysis considers thermal symmetric dark matter models where 
self-interactions are mediated by a dark vector boson which couples to the SM via mass 
mixing with the $Z$ boson and kinetic mixing. We consistently considered all the relevant 
phenomenology: dark matter self-interactions, enhanced annihilation due to the Sommerfeld 
effect, momentum suppression for dark matter--nucleon scatterings, and the possibility of 
enhanced neutrino signal due to the mediator late decays \cite{Bell:2011sn}. 
This latter effect was very recently studied focusing mainly on the
possibility of new gamma ray and charged lepton signatures of dark matter 
coupled to the SM trough dark photons \cite{Feng:2016ijc} and collisionless dark matter 
\cite{Leane:2017vag}.
Additionally, we considered a few combinations for the mixing parameters, leading to different dark 
matter couplings with neutrons and protons. We also compare our results with those from direct detection analysis.  

We organize this paper as follows: section \ref{sec:vdSIDM} briefly 
describes the vdSIDM model assumed in this analysis, and summarize current 
constraints on the model parameters. Section \ref{sec:DM_cap&ann} describes how we 
estimate the dark matter capture and annihilation rates in the Sun. This is followed 
by our predictions for the signal and background events in section \ref{sec:Nu}. The 
discussion of our results follows in section \ref{sec:sens}, and we conclude in the 
last section. 

\section{Velocity Dependent SIDM model} \label{sec:vdSIDM}


We consider the  dark matter particle $\chi$ as a Dirac fermion
which couples to a vector mediator $\phi_\mu$ of a dark $U(1)_\chi$ gauge interaction 
through the Lagrangian term:
\begin{equation} \label{eq:L_int}
  \mathcal{L} = g_\chi \bar{\chi} \gamma^\mu \chi \phi_\mu,
\end{equation}
where $g_\chi$ is the coupling constant. This interaction gives rise to dark matter 
self-scatterings and annihilations. We also take the dark matter as symmetric, with 
equal abundance of  particles and anti-particles.


In the non-relativistic limit, dark matter self-interactions can be described by a 
Yukawa potential \cite{Feng:2009hw, Loeb:2010gj, Buckley:2009in, Aarssen:2012fx}
\begin{equation} \label{eq:Yuk}
  V(r) = \pm \frac{\alpha_\chi}{r} e^{-m_\phi r},
\end{equation}
where $\alpha_\chi = g_\chi^2/(4 \pi)$ is the dark fine structure constant, $m_\phi$ 
the mediator mass and $r$ the relative distance between the dark
matter particles.
This potential is attractive ($-$) for $\chi \bar{\chi}$ scatterings, 
and repulsive $(+)$, for $\chi \chi$ and $\bar{\chi} \bar{\chi}$. 

\begin{figure}[!hbtp]
  \begin{center}
    \includegraphics[width=0.8\textwidth]{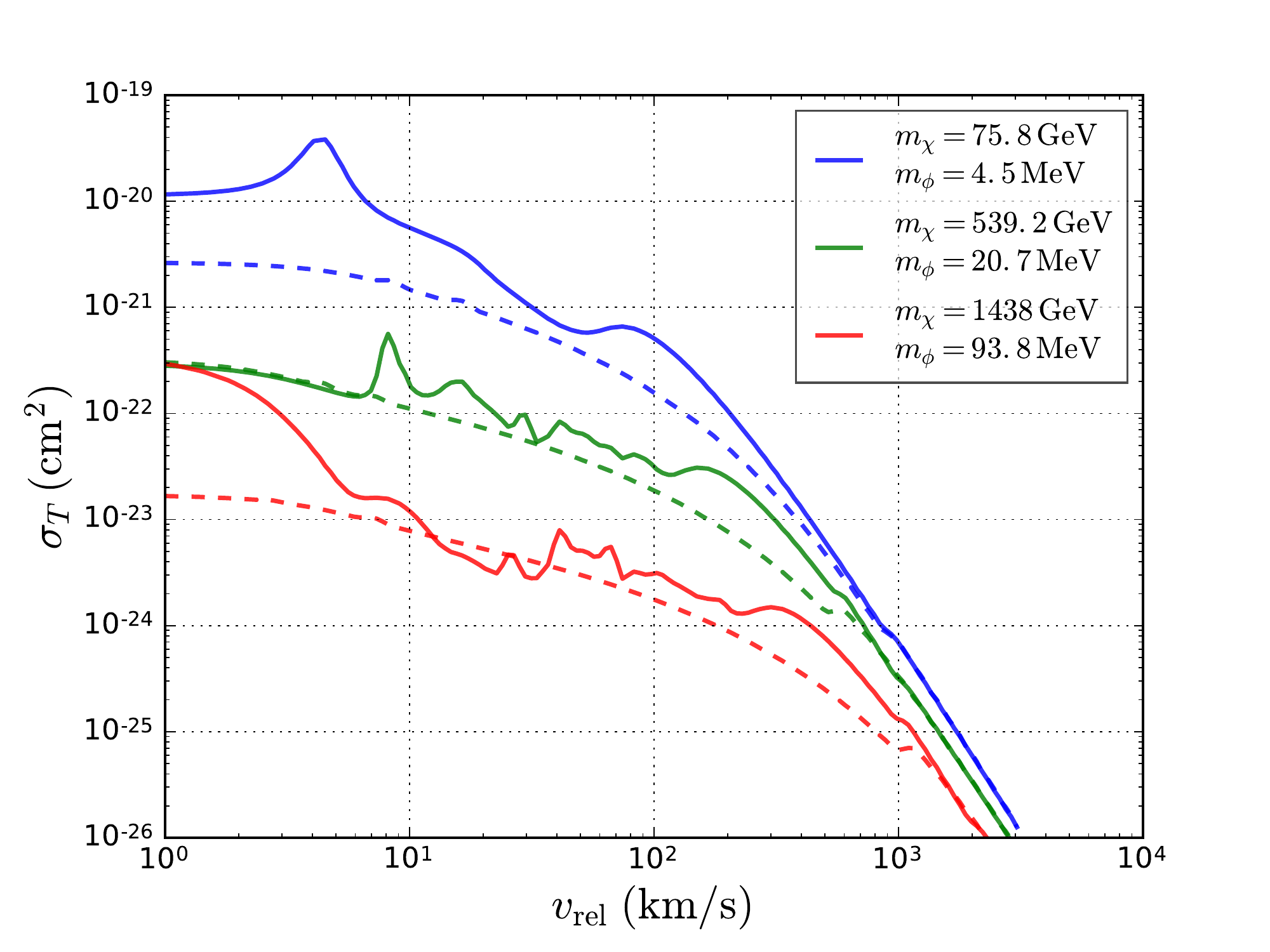}
  \end{center}
  \caption{Self-scattering dark matter cross section as a function of the relative
  velocity. Attractive potential (solid curves) and repulsive potentials (dashed curves) 
  are shown for different $m_\chi$ and $m_\phi$ values.
  \label{fig:sigmaT_vrel}}
\end{figure}

The self-scattering cross section is determined \cite{Tulin:2013teo} by numerically
solving the Schrödinger equation with the Yukawa potential.
A partial wave analysis is used when the classical (Born) limit breaks down at 
$m_\chi v/m_\phi \lesssim 1$ ($\alpha_\chi m_\chi/m_\phi \gtrsim 1$). Figure \ref{fig:sigmaT_vrel}  
shows the self-scattering transfer cross section \footnote{In common with the literature, we used $\sigma_T$ as an effective 
cross section that relates the dark matter particle physics to the results from simulations 
of structure evolution. For details see \cite{Tulin:2013teo}.}, 
$\sigma_T = \int d\Omega (1 -\cos \theta) d\sigma_{\chi \chi}/d\Omega$,
as a function of the relative velocity, for both attractive (solid curves) and repulsive 
potentials (dashed curves). The strong dependence of the self-scattering cross section on 
the velocity is evident, decreasing several orders of magnitude as the velocity goes from $v=10$ to 1000 km/s. 
Also, for the attractive case, the velocity dependence is not trivial, presenting several 
resonances.

The dark matter annihilation cross section at tree level is given by
\begin{equation}
  (\sigma_a v)^\mathrm{tree} = \frac{\pi \alpha_\chi^2}{m_\chi^2} \sqrt{1 - 
    \left( \frac{m_\phi}{m_\chi} \right)^2}.
\end{equation}
However, for low relative velocities, it can greatly increase 
due to the Sommerfeld effect \cite{Somm:1931}. This is 
caused by the attractive self-interactions that distort the wave function of the 
incoming dark matter particles increasing their annihilation
probability. 
The annihilation cross section  can then be represented by
\begin{equation} \label{eq:sigma_av}
  \sigma_a v = S(v) \times (\sigma_a v)^\mathrm{tree},
\end{equation}
where the Sommerfeld factor $S(v)$ can be computed numerically in an 
analogous way as for the self-scattering cross section \cite{ArkaniHamed:2008qn,Hisano:2004ds} 
or analytically by approximating the Yukawa potential by the Hulthen potential 
\cite{Cassel:2009wt,Slatyer:2009vg,Feng:2010zp}.
Using the latter, the Sommerfeld factor is given by 
\begin{equation}\label{eq:Somm}
  S = \frac{\pi}{a} \frac{\sinh(2 \pi ac)}{\cosh(2\pi ac)- \cos (2\pi \sqrt{c - (ac)^2})},
\end{equation}
where $a=v/(2 \alpha_\chi)$ and $c = 6 \alpha_\chi m_\chi/(\pi^2 m_\phi)$.


Additionally, we consider that the dark matter annihilation process 
$\chi \bar{\chi} \rightarrow \phi \phi$ sets the dark matter relic density by thermal 
freeze-out. This requirement fixes the value of $\alpha_\chi$ for given values of 
$(m_\chi,m_\phi)$. This relic density is determined taking the Sommerfeld enhancement into 
consideration \cite{Tulin:2013teo, Feng:2010zp}.  
This effect turns out important only for heavy dark matter $m_\chi \gtrsim 1$ TeV.

\subsection{Light Mediator Couplings to the SM} 


We assume that the dark sector couples to the SM 
via the $\phi$ mediator, which allows production of known particles from dark matter 
annihilation in the Sun. Following a phenomenological approach, 
we consider that the $\phi$ mediator mixes with the photon $\gamma$ and the $Z$ boson 
through
\begin{equation}
  \mathcal{L}_\text{mixing} = \frac{\varepsilon_\gamma}{2} \phi_{\mu \nu} F^{\mu \nu}
    + \varepsilon_Z m_Z^2 \phi_\mu Z^\mu,
\end{equation}
where $\phi_{\mu \nu} \equiv \partial_\mu \phi_\nu - \partial_\nu \phi_\mu$ and 
$F^{\mu \nu}$ are, respectively, the mediator and the photon field strengths.
The first term corresponds to the photon kinetic mixing while the second one to the mass 
mixing with the $Z$\footnote{Although the latter term explicitly breaks gauge
invariance, it can arise in extensions of the SM through the insertion of additional Higgs 
and dark Higgs, as in Ref.\cite{Davoudiasl:2012ag}. In our work, we do not go into the 
details of the dark matter particle physics models, but take a phenomenological approach 
considering both mixing parameters $\varepsilon_\gamma, \varepsilon_Z$ independently. }. 
We take the limit of very small mixing parameters $\varepsilon_\gamma, \varepsilon_Z \ll 1$. 
Both terms are relevant in our analysis, since the $Z$ mixing 
allows the production of high energy neutrinos through $\phi$ decays, and 
the kinetic mixing the dark matter scattering off protons, which
contributes significantly to its capture in the Sun.
Both mixings have been widely studied within the context of vector portal dark matter, dark photon and dark $Z$ 
searches \cite{Holdom:1985ag, ArkaniHamed:2008qn, Pospelov:2008jd, Bjorken:2009mm, Foot:2014uba,
Babu:1997st, Davoudiasl:2012ag, Cui:2017juz, Langacker:2008yv, Cirelli:2016rnw}
and also for vdSIDM direct detection \cite{Kaplinghat:2013yxa}.
 

If only kinetic mixing is present, the $\phi$ decays predominantly
into $e^+e^-$, with decay rate 
\begin{equation} \label{eq:DecWidth_g}
  \Gamma^\gamma_\phi = \frac{\alpha_\text{em} m_\phi \varepsilon_\gamma^2}{3}.
\end{equation}
In the case of $Z$ mixing, the total decay rate is given by
\begin{equation} \label{eq:DecWidth_Z}
  \Gamma^Z_\phi = \frac{\alpha_\text{em} m_\phi \varepsilon_Z^2 (1-\sin^2 \theta_W + 2 \sin^4 \theta_W)}
    {6 \sin^2 \theta_W \cos^2 \theta_W}.
\end{equation}
where the neutrino channel dominates with a branching ratio 
$BR_{(\phi \rightarrow \nu \bar{\nu})} \approx 6/7$, leaving a 
$BR_{(\phi \rightarrow e^+e^-)} \approx 1/7$ for the $e^+e^-$ channel.
Therefore, for given values of $\varepsilon_\gamma$ and $\varepsilon_Z$, the total 
decay rate is $\Gamma_\phi = \Gamma^\gamma_\phi + \Gamma^Z_\phi$ and the branching 
ratio to neutrinos 
$BR_{(\phi \rightarrow \nu \bar{\nu})} = \Gamma^Z_{(\phi \rightarrow \nu \bar{\nu})}/\Gamma_\phi$.


An additional consequence of the mediator mixing with SM particles is the DM scattering 
with nucleons via $\phi$ exchange, which are crucial for dark matter capture in the
Sun. This interaction is represented by
\begin{equation}
  \mathcal{L}_\text{int} = e \phi_\mu (\varepsilon_p \bar{p} \gamma^\mu p
    + \varepsilon_n \bar{n} \gamma^\mu n ),
\end{equation}
where $\varepsilon_p, \varepsilon_n$ are the effective coupling to protons and 
neutrons, and parameterized \cite{Kaplinghat:2013yxa} respectively by:
\begin{align}
  \varepsilon_p &= \varepsilon_\gamma + \frac{\varepsilon_Z}{4 \sin \theta_W \cos \theta_W}
    (1 - 4 \sin^2 \theta_W) 
    \approx \varepsilon_\gamma + 0.05 \varepsilon_Z \\
  \varepsilon_n &= - \frac{\varepsilon_Z}{4 \sin \theta_W \cos \theta_W}
    \approx -0.6 \varepsilon_Z.
\end{align}
Thus, in the case of kinetic mixing, the mediator couples only to the
charged proton.
For $Z$ mixing, the mediator couples mainly to neutrons. So, for these
models isospin violation, i.e. different interactions strengths to protons and neutrons,
arises naturally.


The spin-independent dark matter scattering with a nucleus $N$, carrying atomic number $Z$ 
and mass number $A$, and in the zero momentum transfer limit $(q^2 = 0)$ is given by
\begin{equation}
  \sigma_{\chi N}^\mathrm{SI} = \frac{16 \pi \alpha_\mathrm{em} \alpha_\chi \mu_{\chi N}^ 2}{m_\phi^4} 
    \left( \varepsilon_p Z + \varepsilon_n (A-Z) \right)^2,
\end{equation}
where $\mu_{\chi N}$ is the dark matter--nucleus reduced mass. However, the mediator
masses we are exploring are comparable to the transferred momentum 
in the DM scatterings with the Sun's nuclei, which are typically of the order of 
$q \sim 10$ MeV. Therefore, the cross section is momentum dependent and cannot be 
approximated by a contact interaction \cite{Kaplinghat:2013yxa,Foot:2008nw,Fornengo:2011sz}. We take 
this into account by 
considering a suppression factor
\begin{equation} \label{eq:sigmaXN}
  \sigma_{\chi N}^\mathrm{SI}(q^2) = \sigma_{\chi N}^\mathrm{SI}(q^2=0) \times \frac{m_\phi^4}{(m_\phi^2 + q^2)^2},
\end{equation}
where the momentum transfer is given by $q = \sqrt{2 m_N E_R}$ and $E_R \simeq \mu_{\chi N}^2 v^2/m_N$
is the typical nuclear recoil energy. 


Before describing the dark matter accumulation in the Sun and its associated signal
we briefly summarize some relevant constraints on the parameters we just described.

In order not to violate the standard Big Bang nucleosynthesis, a lower bound on the mixing 
parameters $\varepsilon_\gamma, \varepsilon_Z \gtrsim 10^{-10} \times \sqrt{10 \, \mathrm{MeV}/m_\phi}$ 
comes from the requirement that the $\phi$ mediator decays fast
enough, with a lifetime 
$\tau_\phi \lesssim 1$~s \cite{Lin:2011gj}. 
Also, analysis of supernovae cooling 
through mediator emission establish strong constraints on 
the kinetic mixing parameter, excluding $\varepsilon_\gamma \sim 10^{-10} - 10^{-7}$ for 
$m_\phi \sim 1 - 100$ MeV \cite{Dent:2012mx,Dreiner:2013mua,Kazanas:2014mca,Rrapaj:2015wgs}. 
Recent works have revised these analysis including the plasma effects of finite temperature 
and density \cite{Hardy:2016kme,Chang:2016ntp} and specifically \cite{Chang:2016ntp} excludes 
$\varepsilon_\gamma \gtrsim 10^{-8}$ for $m_\phi \sim 10 - 40$ MeV and 
$\varepsilon_\gamma \gtrsim 10^{-9}$ for $m_\phi \sim 10$ MeV independently of the details
in their modeling.

Other constraints come from beam dump and fixed target experiments such as
SLAC E137 \cite{Bjorken:1988as,Bjorken:2009mm}, the LSND neutrino experiment 
\cite{Batell:2009di,Essig:2010gu} and CHARM \cite{Bergsma:1985is,Gninenko:2012eq}.
Among these, the strongest results correspond to E137, excluding 
$\varepsilon_\gamma \gtrsim 10^{-7}$ for mediator masses $m_\phi \lesssim 400$ MeV.

Additional constraints come from dark matter direct detection searches. Recent analysis
have used the results from XENON100, LUX and SuperCDMS experiments to constrain 
vdSIDM models \cite{Kaplinghat:2013yxa,DelNobile:2015uua}. Their results indicate that most 
of the relevant parameter space with $m_\chi \gtrsim 10$~GeV if $\varepsilon_\gamma, \varepsilon_Z = 10^{-8}$ 
is excluded, and $m_\chi \gtrsim 30$~GeV if $\varepsilon_\gamma, \varepsilon_Z = 10^{-9}$. 

Finally, there are also strong constraints from indirect detection searches. Recent
analysis that included the enhancement in dark matter annihilations due to the Sommerfeld
effect \cite{Bringmann:2016din,Cirelli:2016rnw} have found that the observations of gamma 
rays by Fermi-LAT \cite{Ackermann:2015zua}, the positron and anti-proton flux by AMS-02 
\cite{Accardo:2014lma,Aguilar:2014mma,Aguilar:2016kjl} and particularly the CMB by Planck 
\cite{Ade:2015xua} exclude a large part of
the parameter space of vdSIDM that can alleviate the CDM small scale problems. 
However, models with $m_\phi \lesssim 1$ MeV still remain a possible solution 
in the case of $\phi-Z$ mass mixing.

In our analysis we focus on vdSIDM models with mixing parameters between 
$\varepsilon_\gamma, \varepsilon_Z \sim 10^{-10} - 10^{-8}$ and with $m_\chi \sim 1$ GeV - TeV 
and $m_\phi \sim 1 - 100$ MeV.

\section{Dark Matter Capture and Annihilation in the Sun} \label{sec:DM_cap&ann}


In this section we analyze how dark matter self-scattering affects the capture and 
annihilation processes in the Sun. The time evolution for the total
number of dark matter particles and anti-particles\footnote{From here on when we mention dark matter particles,
  we consider them in conjunction with their anti-particles, unless specified.} in the Sun
$N_\chi$ is 
\begin{equation}\label{eq:Nx_evol}
  \dot{N_\chi} = \Gamma_c + \Gamma_s - 2 \Gamma_a,
\end{equation}
where $\Gamma_a$ is the dark matter annihilation rate, $\Gamma_c$ its capture rate 
due to scatterings with the Sun's nuclei, and $\Gamma_s$ its self-capture rate. Note 
that evaporation processes were neglected, given that they are
negligible for DM masses $m_\chi \geq 4$ GeV \cite{Gould:1987ju,Griest:1986yu}.
Neither the effects of dark matter bound states formation are
considered, since these
are expected to be negligible for most of the parameter space that we study 
\cite{vonHarling:2014kha,Cirelli:2016rnw}\footnote{
The formation and decay of dark matter bound states in the Sun can contribute 
positively to dark matter indirect signals for large dark matter masses ($m_\chi \gtrsim 2$ TeV)
\cite{Cirelli:2016rnw}, making our results for neutrino telescopes' sensitivities 
conservative at these $m_\chi$ values.}.


The DM capture rate due to scatterings with the Sun's nuclei is given by
\begin{equation}
  \Gamma_c = 4.8\times10^{24} \, \text{s}^{-1} \frac{\rho_\chi}{0.3 \, \text{GeV}/\text{cm}^3} 
	\left( \frac{270 \, \text{km}/\text{s}}{\bar{v}} \right) 
	\left( \frac{\text{GeV}}{m_\chi} \right) \sum_i \mathcal{F}(m_\chi,N_i)
	\left( \frac{\sigma_{\chi N_i}}{10^{-40} \, \text{cm}^2} \right),
\end{equation}
where we take the local dark matter halo density $\rho_\chi = 0.4$ GeV/cm$^3$ 
\cite{Salucci:2010qr,Catena:2009mf}, and its velocity dispersion $\bar{v} = 270$ km/s. 
The sum is over all nuclear species $N_i$ in the Sun and the factor $\mathcal{F}(m_\chi,N_i)$ denotes 
the product of several factors for each nuclear element. It includes their mass 
fraction, distribution in the Sun, and the kinematic and form factor suppressions 
\cite{Jungman:1995df}.
For the first two of these factors, we took updated values from \cite{Asplund:2009fu} 
and \cite{Bahcall:2004pz}. The dark matter--nucleus cross section 
$\sigma_{\chi N_i}$ is given by equation (\ref{eq:sigmaXN}).


The dark matter capture rate has been determined in 
\cite{Press:1985ug,Gould:1987ir,Gould:1991hx}, and extended in \cite{Zentner:2009is,Catena:2016ckl} 
to include self-interactions. The SIDM high energy neutrino rate in IceCube has been 
determined in \cite{Albuquerque:2013xna}. Here we extend these
procedures to account for the velocity dependence of the self-scattering cross 
section $\sigma_{\chi \chi}(v_\mathrm{rel})$, and also the spatial and velocity 
distributions of the dark matter particles in the Sun. 
The self-capture rate can be written as 
$\Gamma_s = C_s N_\chi$,
where $N_\chi$ is the number of captured dark matter particles 
and
\begin{equation} \label{eq:Cs}
  C_s = \int \frac{f_\mathrm{halo}^\mathrm{DM}(u)}{u} \sigma^\mathrm{eff}(v_\mathrm{rel}) f_\mathrm{sun}^\mathrm{DM}(r,u^\prime) 
	  (v^2 - {u^\prime}^2 - u^2) \varTheta(v^2 - {u^\prime}^2 -u^2) 4 \pi r^2 dr d\theta du^\prime du,
\end{equation}
where
\begin{equation}
  f_\mathrm{halo}^\mathrm{DM}(u) = \sqrt{\frac{6}{\pi}} \left( \frac{\rho_\chi}{m_\chi} \right) \frac{1}{\bar{v}}
	  e^{ -\frac{3}{2}(u/\bar{v})^2 } e^{ -\frac{3}{2} (\tilde{v}/\bar{v})^2 }
	  u^2 \sinh( \frac{3u\tilde{v}}{\bar{v}^2} ) \frac{1}{u \tilde{v}}
\end{equation}
is the dark matter velocity distribution in the halo, 
$\tilde{v} = 220$ km/s is the Sun's velocity and $v$ is the escape velocity at 
a distance $r$ from the center of the Sun.
We assume that the dark matter particles are thermally distributed in the Sun 
\cite{Gould:1987ju}:
\begin{equation}
  f_\mathrm{sun}^\mathrm{DM}(r,u^\prime) = \frac{1}{V_1} \left( \frac{m_\chi}{2\pi k T_\chi} \right)^{3/2}
	  e^{ -m_\chi {u^\prime}^2 /( 2kT_\chi ) } e^{ -m_\chi \phi(r)/( kT_\chi ) }
	  2 \pi {u^\prime}^2 \sin \theta,
\end{equation}
where $V_1 = \int_0^{R_\odot} e^{- m_\chi \phi/(kT_\chi)} 4\pi r^2 dr$, $\phi(r)$ is the
solar gravitational potential, 
and $T_\chi$ is the dark matter temperature 
in the Sun\footnote{For most $m_\chi$ values, the dark matter temperature is equal to 
the Sun's core temperature $T_\chi = 1.57 \times 10^7$~K. However for 
$m_\chi \lesssim 10$ GeV a correction is needed, since in this case the particles are 
more dispersedly distributed around the solar core, reaching distances where the 
solar temperature is lower. Hence, for these lower masses, we considered $T_\chi$
equal to the solar temperature at the dark matter mean distance from the Sun's core,
e.g. $T_\chi = 1.43 \times 10^7$ K and $1.52 \times 10^7$ K for $m_\chi = 4$ and 10 GeV 
respectively.}. The effective dark matter self-scattering cross section is 
$\sigma^\mathrm{eff}(v_\mathrm{rel}) = (\sigma^\mathrm{att}_{\chi \bar{\chi}} + \sigma^\mathrm{rep}_{\chi \chi})/2$, 
accounting for both  
attractive and repulsive interactions. The relative speed is 
$v_\mathrm{rel} = ( v^2 + u^2 + {u^\prime}^2 - 2 u^\prime \sqrt{ v^2 +
  u^2 } \cos \theta )^{1/2}$, where 
$\theta$ the angle between the velocity of the incoming particle and the one already 
captured in the Sun. Finally, $\varTheta$ is the Heaviside step function. We integrate 
equation(\ref{eq:Cs}) over the dark matter particles' velocities in the halo  
$(u)$ and in the Sun $(u^\prime$) and over the Sun's volume.

\begin{figure}[!hbtp]
  \begin{center}
    \includegraphics[width=0.9\textwidth]{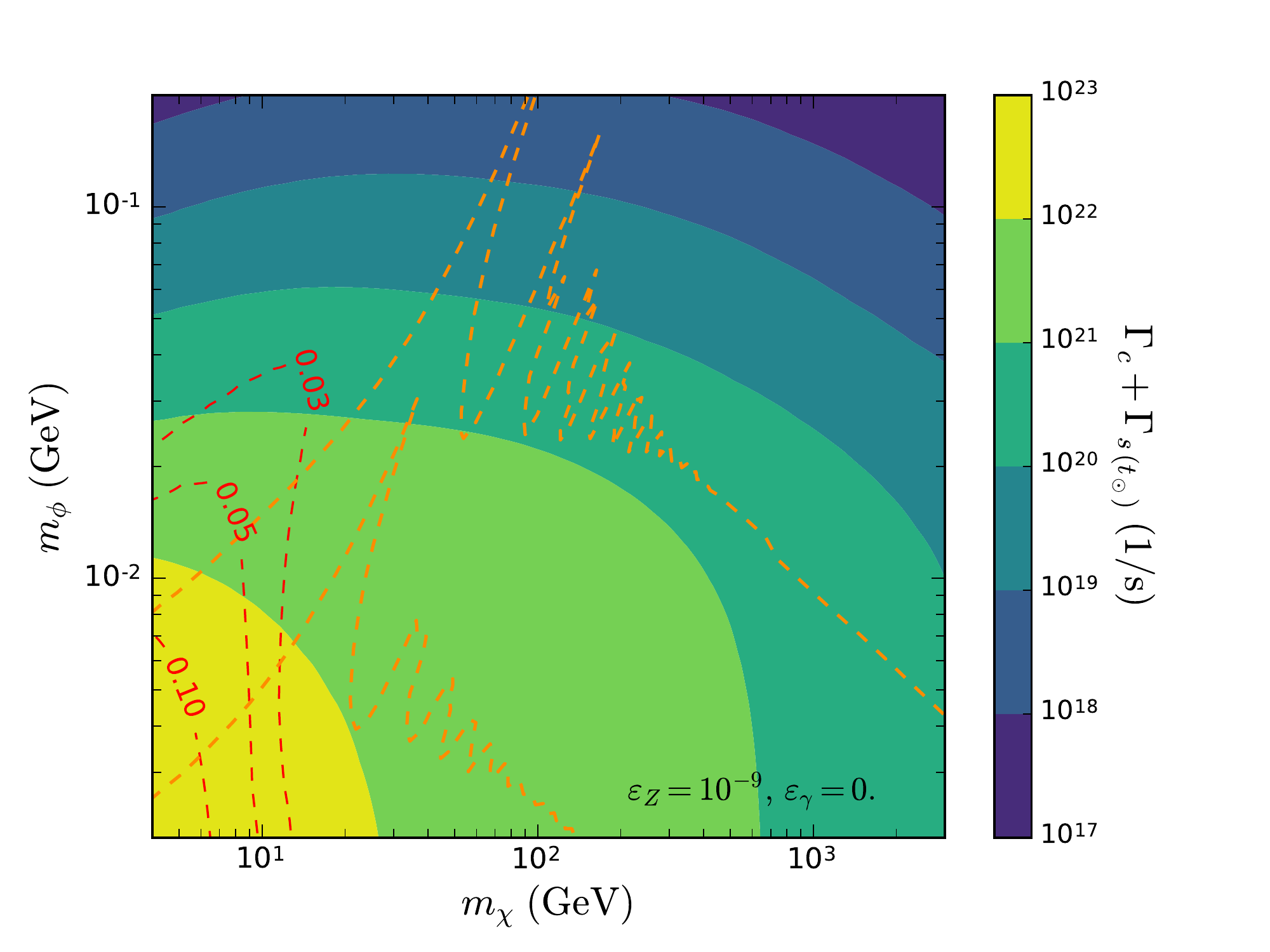}
  \end{center}
  \caption{Total dark matter capture rate, where $\varepsilon_Z = 10^{-9}$ and $\varepsilon_\gamma = 0$.
  \label{fig:totCapRate}}
\end{figure}

Figure \ref{fig:totCapRate} shows the total dark matter capture rate
$\Gamma_c + \Gamma_s(t_\odot)$ at the present epoch $t=t_\odot$ for 
$\varepsilon_Z = 10^{-9}$ and $\varepsilon_\gamma = 0$. The red dashed contours 
indicate the self-capture contribution 
relative to the total rate, while the region in between the orange dashed 
curves indicates the vdSIDM region that alleviates the too big to fail problem with 
$0.1 < \langle \sigma_T \rangle /m_\chi <10$ cm$^2$/g, assuming a 
characteristic dwarf velocity of $v_0 = 30$ km/s.
It is clear that the dark matter self-interaction contribution to the total capture 
rate is negligible for most of the parameter space.

\begin{figure}[!hbtp]
  \begin{center}
    \includegraphics[width=0.9\textwidth]{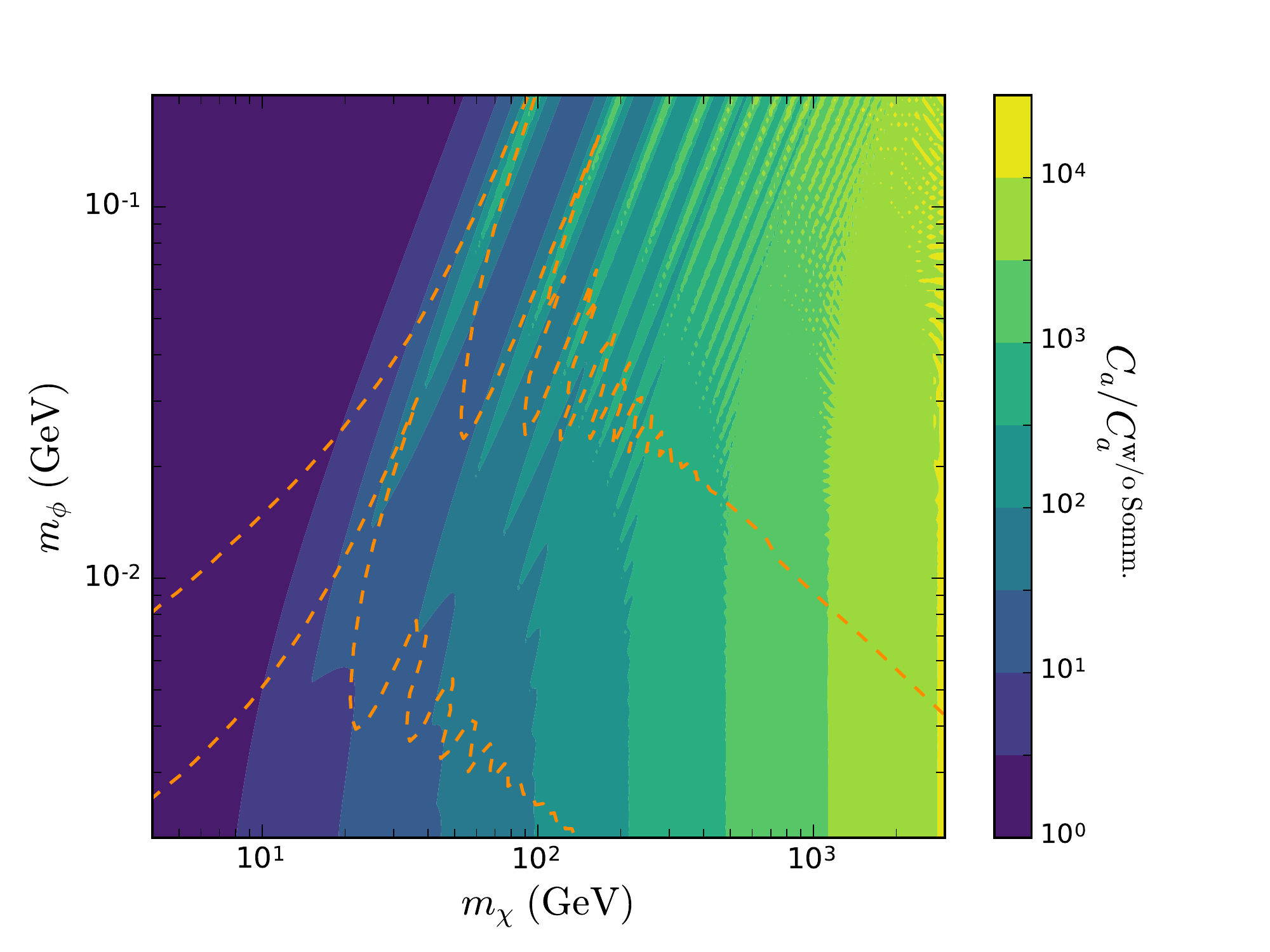}
  \end{center}
  \caption{Sommerfeld enhancement effect on $C_a$.
  \label{fig:Ca}}
\end{figure}

The annihilation rate $\Gamma_a$ is given by
\begin{equation}
  \Gamma_a = \int d^3r \, n_\chi(r) n_{\bar{\chi}}(r) \langle \sigma_a v \rangle,
\end{equation}
where $\langle \sigma_a v \rangle$ is the thermally averaged dark matter annihilation
cross section and $n_\chi(r)$ 
is the radial distribution of dark matter particles 
in the Sun. This equation can be rewritten as $\Gamma_a = C_a N_\chi^2/4$, where 
$N_\chi$ is the total number of dark matter particles 
and $C_a = \langle \sigma_a v \rangle/ V_\mathrm{eff}$, with effective volume
$V_\mathrm{eff} = 6.9 \times 10^{27} ( \frac{T_\chi}{1.57 \times 10^7 ~\mathrm{K}} )^{3/2}
  ( \frac{100 ~\mathrm{GeV}}{m_\chi} )^{3/2} ~\mathrm{cm}^3$ \cite{Griest:1986yu}.
The thermally averaged cross section 
is
\begin{equation}
  \langle \sigma_a v \rangle = 
    \left( \frac{m_\chi}{4\pi k T_\chi} \right)^{3/2} \int d^3 v \, e^{-\frac{m_\chi v^2}{4k T_\chi}} (\sigma_a v),
\end{equation}
where $(\sigma_a v)$ includes the Sommerfeld enhancement factor. The effect of 
this factor over $C_a$ is clearly seen in figure \ref{fig:Ca}, 
where the ratio $C_a/C_a^\mathrm{w/o \; Somm.}$ of $C_a$ including over excluding the
Sommerfeld factor is shown. It can be seen that $C_a$ is greatly enhanced by this 
factor, specially at large masses $m_\chi > 100$ GeV. This is expected, since the 
average dark matter velocity in the Sun decreases for large $m_\chi$.

Once the total capture and annihilation rates are obtained,
the total number of DM particles in the Sun at a age $t$ is given by 
equation(\ref{eq:Nx_evol}), which  yields to
\begin{equation}
  N_\chi(t) = \frac{\Gamma_c \tanh(t/\zeta)}{\zeta^{-1} - C_s \tanh(t/\zeta)/2},
\end{equation}
where $\zeta = (\Gamma_c C_a/2 + C_s^2/4)^{-1/2}$
\cite{Zentner:2009is}. 
So, at present time $t_\odot = 4.57 \times 10^9$ years, and the annihilation rate is 
$\Gamma_a = C_a N_\chi^2(t_\odot)/4$. 

\begin{figure}[!hbtp]
  \begin{center}
    \includegraphics[width=0.49\textwidth]{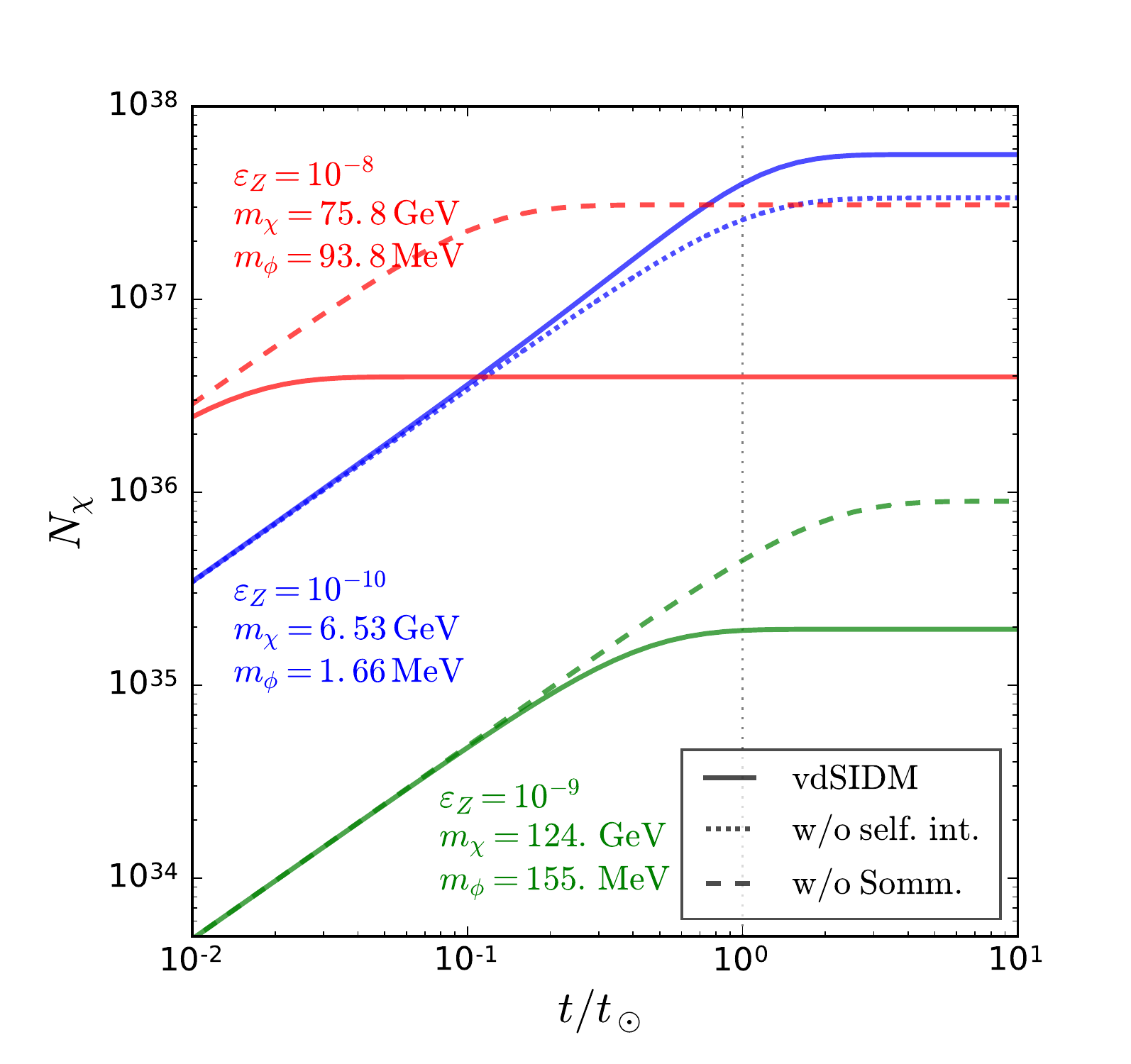}
    \includegraphics[width=0.49\textwidth]{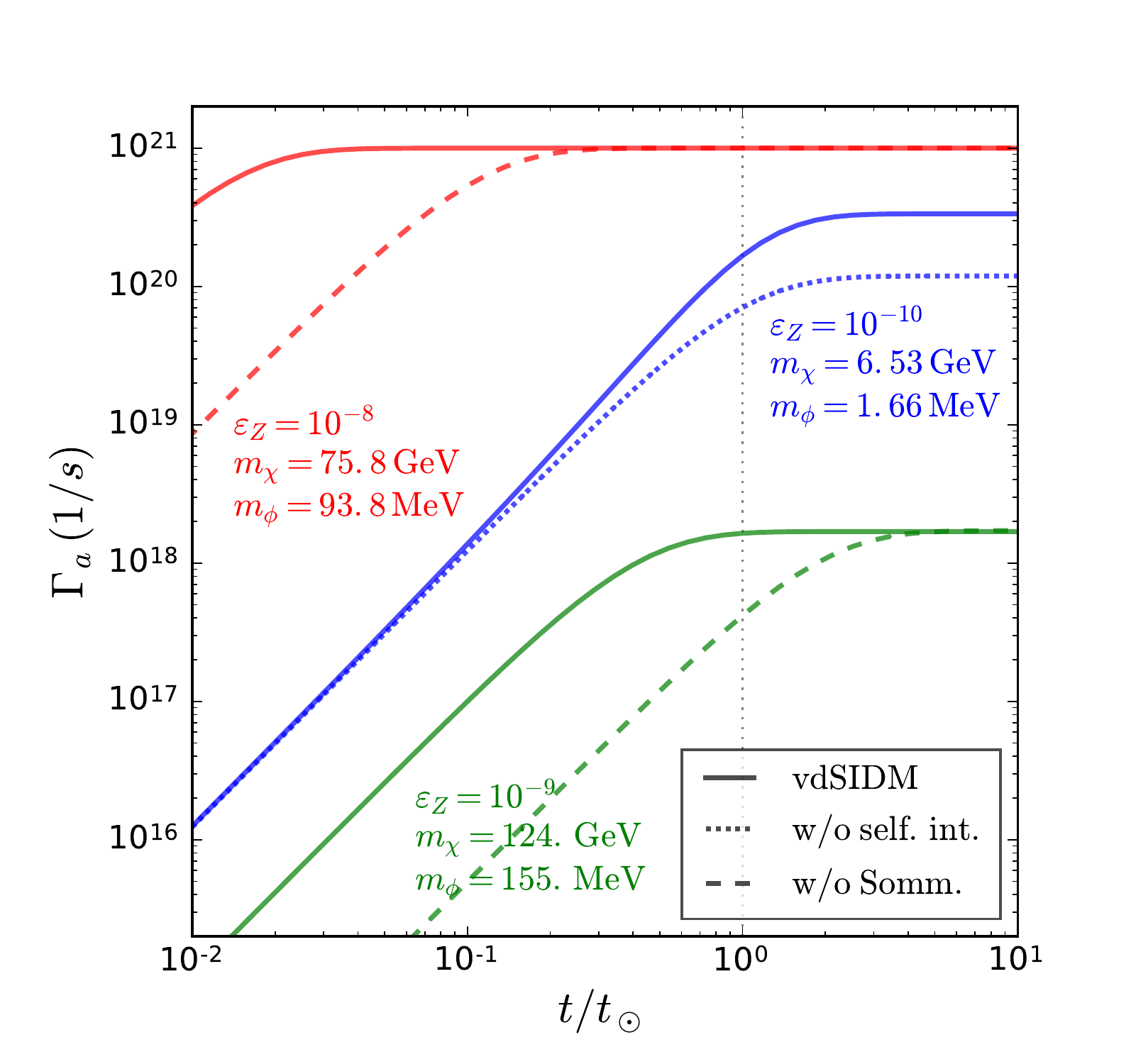}
  \end{center}
  \caption{The time evolution of captured dark matter particles for three 
  representative cases (left) and their corresponding annihilation rates (right). 
  Solid curves represent results for the full calculation
  as described in the text, dotted curves neglect the DM self-capture term, dashes 
  curves ignore the Sommerfeld enhancement in DM annihilations.
\label{fig:Nx_Ga_vs_t}}
\end{figure}

The time evolution of the captured dark matter particles 
and their corresponding annihilation rates 
are shown in figure \ref{fig:Nx_Ga_vs_t}. The solid curves represent our results for
the full calculation 
as described above, while the dotted curves neglect the DM self-capture term, and the 
dashed curves ignore the Sommerfeld enhancement in DM annihilations. 
The effect of DM self-interactions is only noticeable when the DM--nucleon cross 
section is very low, which happens for $\varepsilon_Z = 10^{-10}$ (blue curves). In 
this case, 
both the number of captured particles and the annihilation rate increase.
On the other hand, the Sommerfeld enhancement in the annihilation rate hastens 
the equilibrium. This effect is more evident for $\varepsilon_Z = 10^{-8}$ and $10^{-9}$ 
(red and green curves). Although this lowers the total number of captured particles, 
the annihilation rate is much larger and reaches its maximum much earlier.
However, notice that the Sommerfeld enhancement does not necessarily 
cause a larger annihilation rate at the present time (highlighted by the dotted vertical 
line), as is the case for $\varepsilon_Z = 10^{-8}$.

\section{Neutrino Signal and Background} \label{sec:Nu}


The annihilation of dark matter particles in the Sun creates pairs of $\phi$ particles
which, through their subsequent decay, 
produces high energy electrons and neutrinos. In this analysis we focus on the neutrino signal, 
estimating its flux at Earth and both IceCube-DeepCore \cite{Collaboration:2011ym} 
and PINGU's \cite{Aartsen:2014oha} sensitivity to vdSIDM models. In this way, we 
determine the vdSIDM parameter space to be probed by these telescopes.


The neutrinos and anti-neutrinos\footnote{From here on when we mention 
neutrinos, we consider them in conjunction with anti-neutrinos, unless specified.}
production is flavor blind, and their 
energy spectrum per annihilation is given by:
\begin{equation}
  \frac{dN_\nu}{dE_\nu} = \frac{4}{\Delta E} \varTheta(E_\nu - E_-) \varTheta(E_+ - E_\nu),
\end{equation}
where $\Delta E = \sqrt{m_\chi^2 - m_\phi^2}$ with its maximum and minimum 
energies at $E_\pm = (m_\chi \pm \sqrt{m_\chi^2 - m_\phi^2})/2$, 
or in our case $E_- \approx 0$ and $E_+ \approx m_\chi$.  
Neutrino production is maximum when only $Z$ mass mixing is considered, or in other 
words, when $\varepsilon_\gamma = 0$. 
Here we consider this scenario, for which $BR(\phi \rightarrow \nu \bar{\nu}) \approx 86\%$. 
We also consider additional cases, where 
$\varepsilon_\gamma = -0.64 \times \varepsilon_Z$, such that the $\phi$ couplings to 
protons and neutrons are equal ($\eta \equiv \varepsilon_n/\varepsilon_p = 1$), with
$BR(\phi \rightarrow \nu \bar{\nu}) \approx 74\%$, and $\eta = -0.7$ with a $68\%$ 
branching ratio. 

\begin{figure}[!hbtp]
  \begin{center}
    \includegraphics[width=0.9\textwidth]{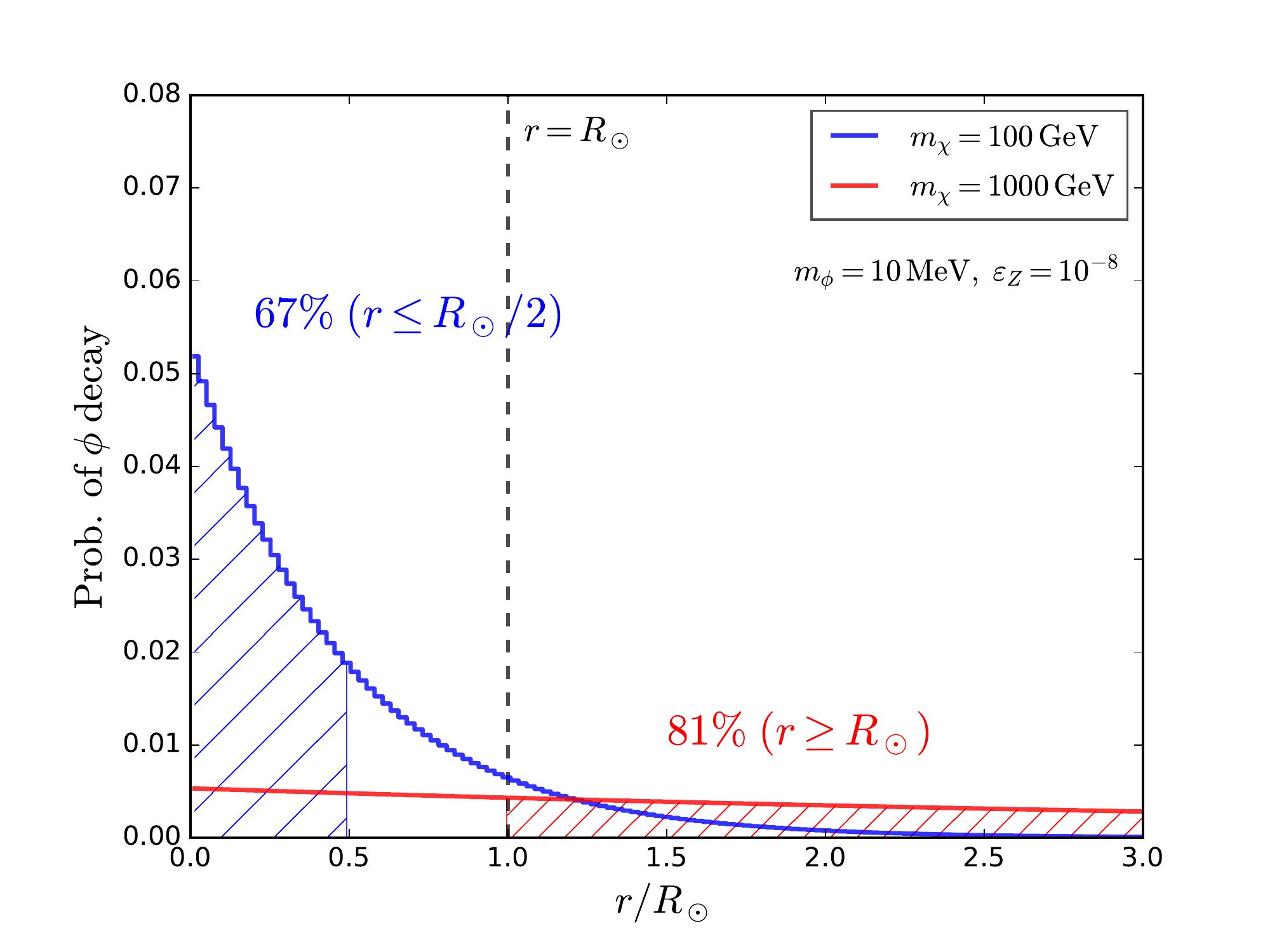}
  \end{center}
  \caption{$\phi$ mediator  decay probability as a function of the
    distance from the Sun's core. 
    For 100 GeV dark matter, most decays ($67\%$) occur inside the Sun with 
    $r \leq R_\odot/2$, while 
    for 1~TeV, most decays ($81\%$) occur outside the Sun.
  \label{fig:pPhi_decay}}
\end{figure}

It is important to note that the neutrinos are not necessarily produced at
the Sun's core, since the $\phi$ mediators propagate freely until
their decay.\footnote{Due to the very small mixing parameters
  $\varepsilon_\gamma, \varepsilon_Z$,  the 
$\phi$ mediator--nucleon cross section
is also very small, being $\sigma_{\phi p} \sim 10^{-46}$ cm$^2$, which implies that 
their interaction length is much greater than the Sun radius.}
Thus, the neutrino production point depends strongly 
on the mixing parameters as well as on the $\phi$'s Lorentz factor.
For example, for pure $Z$ mixing, the mean decay length in the Sun 
is given by 
\begin{equation}
  \lambda_\phi \approx 4 \times 10^{-2} \, R_\odot \times \left( \frac{10^{-8}}{\varepsilon_Z} \right)^2
    \left( \frac{10 \, \mathrm{MeV}}{m_\phi} \right) 
    \left( \frac{m_\chi/m_\phi}{1000} \right)
\end{equation}
The $\phi$ decay probability distribution is shown in figure 
\ref{fig:pPhi_decay} for two cases: $\varepsilon_Z = 10^{-8}, m_\phi = 10~\mathrm{MeV}, 
m_\chi = 10~\mathrm{GeV}$ (blue) and $m_\chi = 1~\mathrm{TeV}$ (red). In the blue
distribution most ($\sim 67\%$) of the $\phi$ mediators decay within the inner part of the
Sun ($r \leq R_\odot/2$), while for the red distribution, due to the larger Lorentz
boost, most decays ($\sim 81\%$) occur outside the Sun.

To determine the neutrino flux at the detector, we developed a 
simulation code where the neutrino point of production is selected
accordingly to the $\phi$ decay distribution. 
For each combination of parameters
($\varepsilon_\gamma, \varepsilon_Z, m_\chi, m_\phi$) we considered $5 \times 10^5$ 
annihilation events. Neutrinos were propagated to the detector 
\cite{Barger:2007xf,Cirelli:2005gh,Blennow:2007tw,Ohlsson:2001et} taking
into account neutral and charged current 
interactions, oscillations and the production of secondary neutrinos.
We used the density profile of the Sun as given by the standard solar 
model BS05OP \cite{Bahcall:2004pz}. Propagation from the Sun's surface to a distance 
of 1 AU (Earth's approximate average distance from the Sun), included neutrinos
produced in this region, as well as oscillations in the
vacuum. Finally, 
they were propagated from 1~AU to the detectors's location at the Earth considering their 
observational time. For this latter step we used the WimpEvent program, 
contained in the WimpSim simulation 
package \cite{Blennow:2007tw}.~\footnote{For consistency we used WimpSim~3.05 
default values for the neutrino oscillation parameters, as well as the 
neutrino--nucleon cross sections (calculated with CTEQ6-DIS PDFs \cite{Pumplin:2002vw}).}


The number of neutrino signal events $N_\nu^\mathrm{s}$ in the detector
is given by
\begin{equation}
  N_\nu^\mathrm{s} = \Gamma_a t_\mathrm{exp} \times \int_{\Delta \Omega} \int_{E_\mathrm{th}}^{m_\chi} 
    \frac{d^2\phi_{\nu}}{dE_\nu d\Omega} A_\mathrm{eff}(E_\nu) dE_\nu \, d\Omega
\end{equation}
where 
$t_\mathrm{exp}$ is the detector's exposure 
time, $\phi_\nu$ is the neutrino flux at the detector per
annihilation, 
$E_\mathrm{th}$ is the detector energy threshold and the maximum
neutrino energy is 
$m_\chi$. $A_\mathrm{eff}(E_\nu)$ is the detector's 
effective area as a function of the neutrino energy $E_\nu$ 
\cite{Aartsen:2016zhm,Clark:2015idz}. The latter accounts for the neutrino-nucleon 
interaction probability and the produced muon energy loss before
detection, as well as for the detector's trigger and selection analysis efficiencies. 

To estimate the number of signal events we took into account the detector's 
angular resolution. 
The neutrino arrival direction 
was smeared following a gaussian distribution with its standard deviation given by 
the detector's angular resolution \cite{Aartsen:2016zhm,Aartsen:2014oha} at the corresponding neutrino energy. 
Only events arriving within a solid angle $\Delta \Omega = 2 \pi (1 - \cos \Psi)$ 
surrounding the direction between the detector and the Sun were accepted. 
The acceptance angle $\Psi$ depends on energy and specific analysis, and will be 
defined in the next section.


Muons and neutrinos produced by cosmic rays' interactions in the Earth's atmosphere 
constitutes the main background in our analysis. As the IceCube collaboration 
achieves an excellent 
atmospheric muon rejection \cite{Aartsen:2016zhm}, 
we consider that the background 
is comprised 
exclusively by the irreducible flux of atmospheric neutrinos. 

The number of background events is given by 
\begin{equation}
  N_\nu^\mathrm{b} = t_\mathrm{exp} \times \int_{E_\mathrm{th}}^{E_\mathrm{max}} 
    \frac{d\phi_{\nu_\mathrm{atm}}}{dE_\nu} A_\nu(E_\nu) dE_\nu \times \Delta \Omega
\end{equation}
where $\phi_{\nu_\mathrm{atm}}$ is the atmospheric neutrino flux at the detector's
location \cite{Honda:2015fha}.

\section{IceCube-DeepCore and PINGU Sensitivity to vdSIDM} \label{sec:sens}

The IceCube neutrino telescope has a good sensitivity for neutrinos arriving from the 
Sun's direction, and 
thanks to its more recent inner array DeepCore, can lower its energy threshold to 
about $\sim 10$~GeV \cite{Collaboration:2011ym}.  
It also has a good angular resolution for muon neutrinos, ranging from $\sim 5^\circ$ 
at 100 GeV to $\sim 1^\circ$ at 1 TeV \cite{Aartsen:2016zhm}.

In order to determine the IceCube-DeepCore sensitivity to vdSIDM, we considered
the same time period as the latest IceCube collaboration's search for 
dark matter annihilating in the Sun \cite{Aartsen:2016zhm}. It spanned a three year
period including the austral winters between May 2011 and May 2014, with a total
lifetime exposure of $t_\mathrm{exp} = 532$ days. To be compatible with 
their procedure, we only considered up-going muon neutrinos. We split the results 
of our simulations into three samples: events with $m_\chi \leq 50$ GeV, for which
dark matter annihilations result mainly in low energies neutrinos, and therefore can 
only be detected by DeepCore; events with $m_\chi \geq 500$~GeV, for which we 
considered the full IceCube's effective area, and finally events with intermediate 
masses where we performed a combined analysis. Accordingly, we take 
the acceptance angles
$\Psi_1 = 10^\circ$ and $\Psi_2 = 2.8^\circ$ as reference values, where the 
first corresponds to DeepCore's angular cut, as defined in their analysis, and the 
second to the their first angular bin \cite{Aartsen:2016zhm}. This latter value 
allows us to take advantage of the better angular resolution at higher energies.

The number of background events was determined by the average 
atmospheric muon neutrino flux from the Sun's direction in the winter
\cite{Honda:2015fha}.  Since during this season the Sun's zenith angle varies 
between $90^\circ$ and $113.5^\circ$ we took only the average within these directions.

Additionally, we determined 
the planned IceCube's extension PINGU \cite{Aartsen:2014oha} sensitivity to vdSIDM. 
PINGU consists of 40 new strings with 60 optical modules each in the DeepCore region 
of the IceCube detector, lowering the energy threshold to just a few GeV. 
As a consequence, vdSIDM can be probed to lower masses, more specifically 
between 4 and 30 GeV. The same procedure as for
the IceCube-DeepCore analysis was followed for the the PINGU detector, for which we 
used the angular resolution given in \cite{Aartsen:2014oha} and the effective
area in \cite{Clark:2015idz}.

\begin{figure}[!hbtp]
  \begin{center}
    \includegraphics[width=0.9\textwidth]{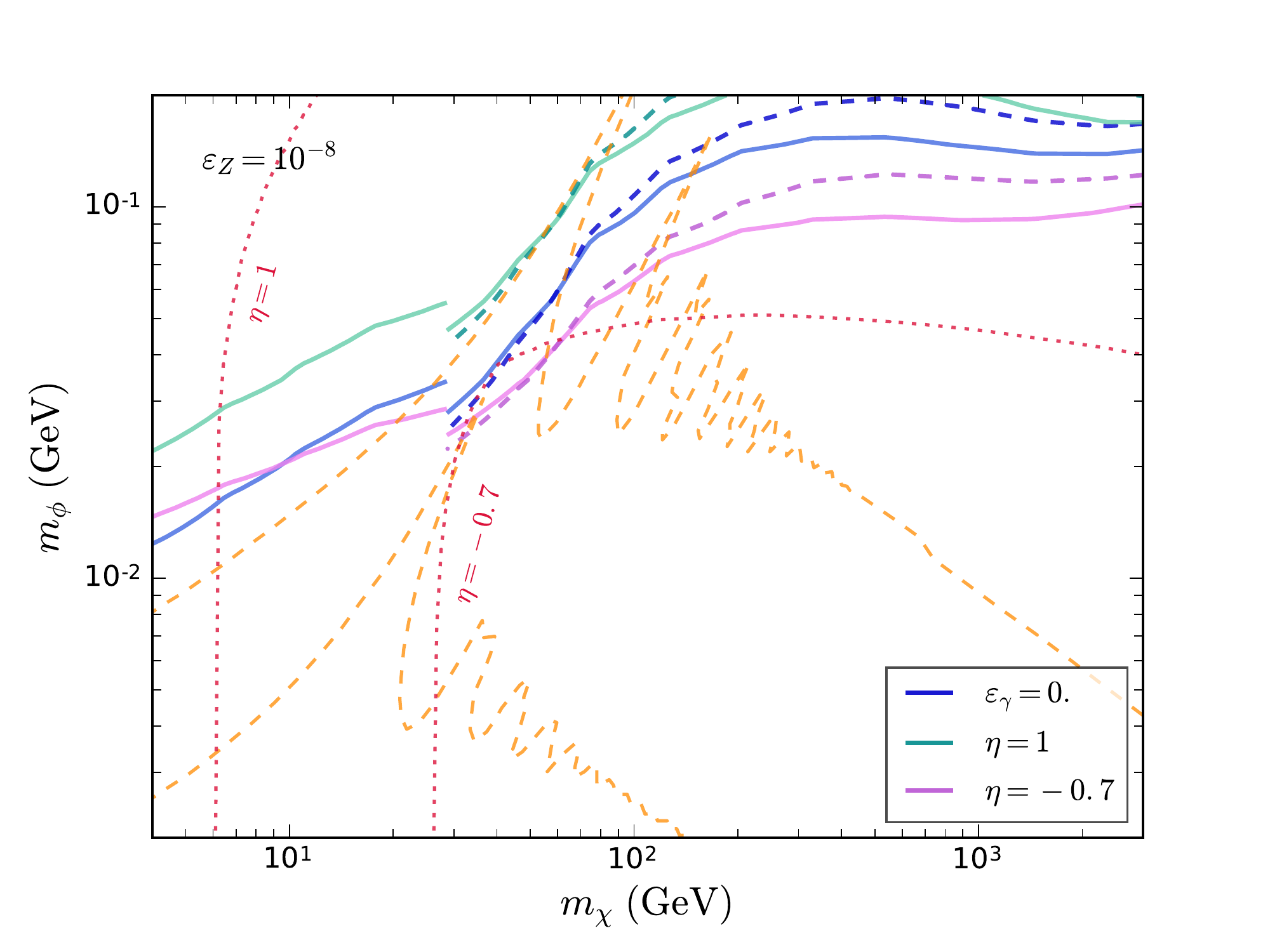}
  \end{center}
  \caption{Sensitivity of IceCube-DeepCore and PINGU for vdSIDM for
    $\varepsilon_Z = 10^{-8}$, where the region below the curves (as labeled) can be 
    probed by these experiments. The region between the orange dashed curves correspond 
    to the vdSIDM parameter space that alleviates the too big to fail problem, having 
    $0.1 < \langle \sigma_T \rangle /m_\chi < 10$ cm$^2$/g. The red dotted curves 
    correspond to limits derived from direct detection (LUX) results, see text for 
    more details.
  \label{fig:sensitivity_eZ-8}}
\end{figure}

\begin{figure}[!hbtp]
  \begin{center}
    \includegraphics[width=0.9\textwidth]{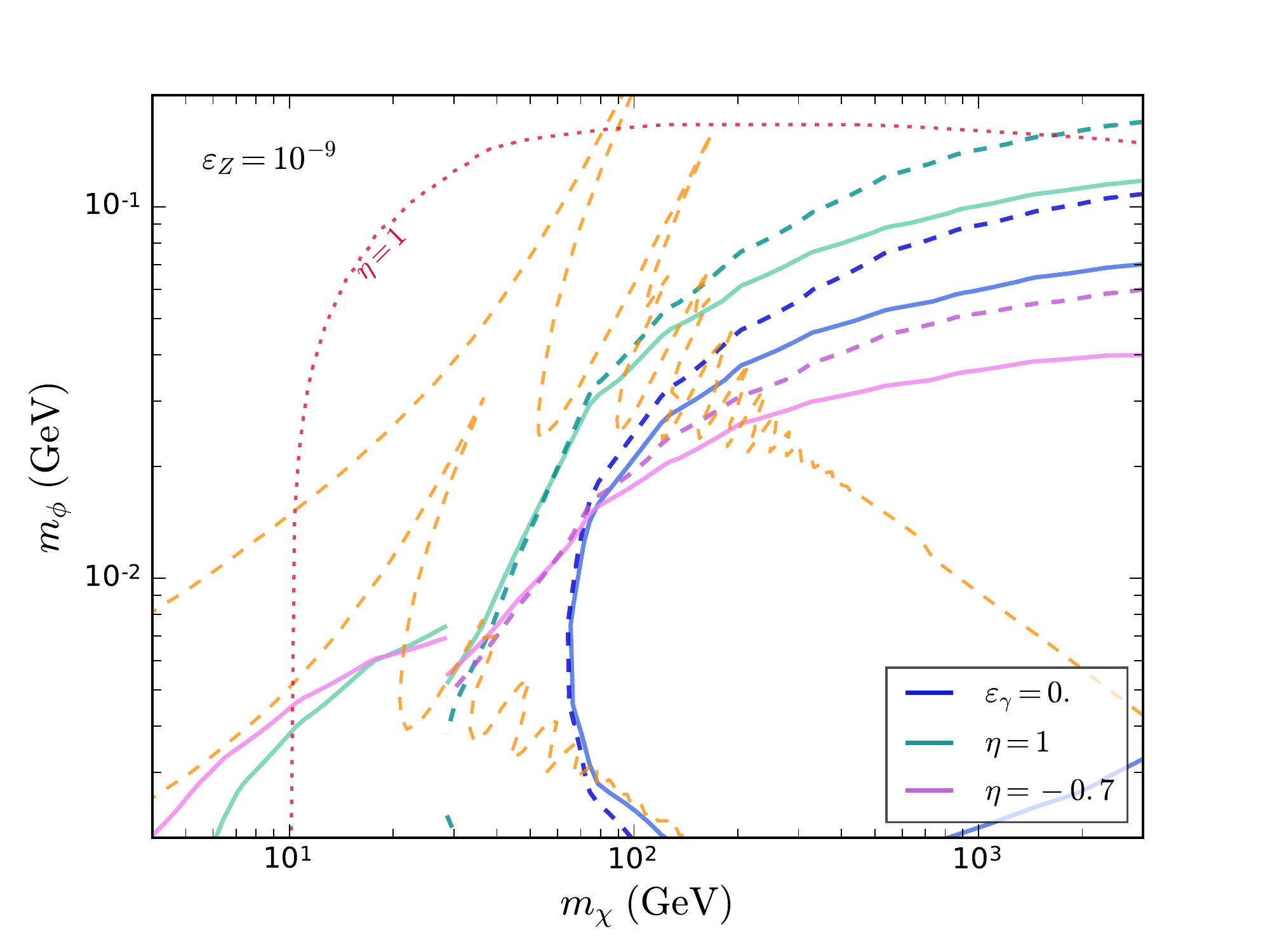}
  \end{center}
  \caption{Same as figure \ref{fig:sensitivity_eZ-8}, but now for $\varepsilon_Z = 10^{-9}$.
  \label{fig:sensitivity_eZ-9}}
\end{figure}

Our results are shown in figures \ref{fig:sensitivity_eZ-8} and \ref{fig:sensitivity_eZ-9},
each for different values of the mixing parameter $\varepsilon_Z$. The regions
below or enclosed by the colored curves 
correspond to $(m_\chi,m_\phi)$ values for which
IceCube-DeepCore and PINGU detectors have at least a 2-sigma detection sensitivity
relative to the expected background of atmospheric muon neutrinos. The solid curves 
are for an acceptance angle $\Psi_1 = 10^\circ$  
while dashed curves are for $\Psi_2 = 2.8^\circ$. Each color correspond to different 
$\varepsilon_\gamma$ or $\eta \equiv \varepsilon_n/\varepsilon_p$ value, as labeled. 
The discontinuity of the curves around $m_\chi = 30$ GeV reflects the 
two individual analysis, one for the IceCube-DeepCore and the other for the PINGU 
detector. The region between the orange dashed curves correspond to the vdSIDM 
parameter space that alleviates the too big to fail problem, having 
$0.1 < \langle \sigma_T \rangle /m_\chi < 10$~cm$^2$/g. For $\varepsilon_Z = 10^{-8}$, 
this parameter space can be almost completely probed by these experiments, 
while for $\varepsilon_Z = 10^{-9}$  a large part of this region can be probed.

For comparison we derive direct detection limits from the LUX experiment 
recent results \cite{Akerib:2016vxi}. We followed the procedure described in 
\cite{Kaplinghat:2013yxa}, considering the $q^2$-dependent suppression factor given 
by equation (\ref{eq:sigmaXN}) and taking $q \approx 50$~MeV for dark matter - xenon 
scattering. Additionally, we determined the limits in the case of isospin violation 
with $\eta = -0.7$ \cite{Feng:2011vu}. These limits are represented by the red dotted 
curves in the figures described above, where the region below them are excluded at 
90$\%$ C.L. Notice that figure \ref{fig:sensitivity_eZ-9} does not include the 
direct detection limit for $\eta = -0.7$ since it falls out of
the explored  parameter space, which indicates the high dependence of direct detection results on the isospin 
violation parameter in contrast to that of the neutrino telescopes.

\section{Conclusions} \label{sec:conc}

We have explored vdSIDM models that can alleviate the small scale structure problems 
observed in dwarf galaxies. We determined neutrino telescopes sensitivity to these models, 
assuming plausible values for its parameters. Our main results are shown in figures
\ref{fig:sensitivity_eZ-8} and \ref{fig:sensitivity_eZ-9}.

We conclude that, for $\varepsilon_Z = 10^{-8}$ the detector's sensitivity is enough to 
probe the vast majority of the relevant vdSIDM parameter space for all 
$\varepsilon_\gamma$ analyzed values, including the low dark matter mass region, thanks to
the PINGU detector. These results show that neutrino telescopes can compete and complement 
the results from dark matter direct detection searches. 

Although the sensitivity decreases for $\varepsilon_Z = 10^{-9}$ it is still 
enough to probe most of the parameter space for 
$m_\chi \gtrsim 70$ GeV. In this case the PINGU detector has almost no sensitivity for 
vdSIDM models. Also, IceCube's sensitivity for low $m_\phi$ and high $m_\chi$ has a 
lower limit, as evidenced by the blue line at the lower right corner of figure 
\ref{fig:sensitivity_eZ-9}. This results from the
$\phi$'s decay length which, in this region, is
larger than the Earth--Sun's distance, and therefore fewer neutrinos are 
produced. Finally, neutrino telescopes lose all sensitivity for $\varepsilon_Z \leq 10^{-10}$. 

We have shown that the IceCube-DeepCore neutrino telescope with its current accumulated 
data is sensitive to most of the parameter space of vdSIDM models that alleviate the small 
scale structure problems observed in dwarf galaxies. Experimental
analysis could probe these models, and independently confirm direct
detection limits. It can also expand these limits, specially in the case of
isospin violation, where we have shown that IceCube-DeepCore sensitivity is drastically
better than direct detection.

\acknowledgments
IA acknowledges the support of the São Paulo Research
Foundation (FAPESP), grant (2016/09084-0). IA also acknowledges the
partial support from the Brazilian National Counsel for
Scientific Research (CNPq). DSR was funded by FAPESP. This project has
received funding from the European Unions Horizon 202 research and
innovative programme under Marie Sklodowska-Curie grant agreement No 674896.
\bibliography{vdsidmRev}{}
\bibliographystyle{JHEP}

\end{document}